# Unresolvable human mental states based on a parallel universe theory

Changsoo Shin[1], Wansoo Ha[1], Wookeen Chung[1], and Sunyoung Park[1]

[1] Department of Energy Systems Engineering, Seoul National University

036-2061, Sillim 9-dong, San 56-1, Gwanak-gu, Seoul, 151-744, SOUTH KOREA

Corresponding Author: Changsoo Shin

Email: css@model.snu.ac.kr

Tel: 82-2-880-7244

Fax: 82-2-875-6296

**Abstract**

We show that human mental states are unresolvable by suggesting a mathematical function that describes human mental states in relation to parallel universe theory. The function is a solution to a multi-dimensional advection equation; representing a situation a person is faced with, and its time-derivative showing the mental state in that situation. This function has interesting characteristics that explain why each person has different thoughts in a particular situation. Because the multi-dimensional advection equation has an infinite number of solutions, we can use them to represent an infinite number of mental states. We focus on the basic concepts of the model and explain the function using extremely simple cases. We also use the functions to explain remembering and forgetting.

## 1. Introduction

Human mental states can be divided into two parts: the conscious and the unconscious. Many psychologists recognized the existence of the unconsciousness in the 20th century. Sidis (1915) used the concept of consciousness and subconsciousness. Freud (1940) divided the human mind into the conscious and the unconscious, and he subdivided the unconscious mind into the preconscious and the unconscious. Jung (1981) divided human mental states into the conscious and unconscious states, and he subdivided the unconscious into personal unconsciousness and collective unconsciousness.

Following Freud and Jung, much research was devoted to analyzing human psychology using various models. A mathematical approach to psychology has emerged since the 1950s (Bush and Mosteller 1951; Estes 1950). Most of the approaches are based on statistical methods and have been applied in many branches of psychology, including models for sensation and perception, learning, memory and thinking, decision making, neural modeling and networks, psychophysics and signal detection, psycholinguistics, motivational dynamics, and psychometric theory (Ashby 1992; Kahneman and Tversky 1979; Krantz 1969; Luce 1997; Regenwetter and Marley 2001; Stevens 1951; Suppes et al. 1989; Townsend and Wenger 2004; Neumann and Morgenstern 1953). However, formulating human mental states using mathematics has been considered impossible due to the inconsistencies, extreme complexities, and irrationalities of the human mind.

In this study, we attempt to show that human mental states are unresolvable by suggesting a simple model in terms of a mathematical function with a parallel universe theory. The function uses the advection equation that formulates human mental states for an extremely simple case. We will focus on an ultimately simple case of the human mind in terms of this function.

## 2. Review of the advection equation

The advection equation (Anderson et al. 1984) was to construct the mental model. One-dimensional advection is expressed as:

$$\frac{\partial U}{\partial t} = -\frac{\partial U}{\partial x} \qquad (1)$$

where $t$ is a time variable and $x$ is a space variable. Equation (1) is equivalent to the one-way wave equation, indicating that time flows in only one direction and, therefore, we can never know the future if we replace term $x$ into another time T. Predicting one's future is impossible because one can change his or her future according to his or her will.

## 3. Human mental state and parallel universes

Schrödinger suggested a thought experiment in which he observed a cat (Fig. 1). According to the parallel universe theory (Everett 1957; Davies 1983), the act of observation makes a universe diverge and both universes, one with a live cat and the other with a dead cat, come into existence. A change in human mental state can be compared with the act of observation; it also makes our universe diverge (Fig. 2) and generates an infinite number of universes. The order of time axes $\tau_k$, $(k = 1,\ 2,\ \cdots,\ n-1)$ in one's universe can be varied and their order affects one's mental state (Fig. 3). Small confusions usually exist in the ordering of time axes in one's memory. Thus, we can have difficulty in ordering past experiences. However, if the confusions are severe, one can suffer from mental disorder. We will discuss one's memory in detail in the following sections.

## 4. Multi-dimensional advection and diverging mental states

We generalized the advection equation for multi-dimensional time as follows:

$$\frac{\partial U}{\partial t} = -\sum_{k=1}^{n} \frac{\partial U}{\partial \tau_k} \quad (2)$$

where $\tau_k$ $(k = 1,2,\cdots,n)$ represents variables of time and we assume that the units of time for both $t$ and $\tau$ are seconds. In the following model, a time axis diverges to multi-time axes every time a change occurs in one's mental state. Because negative or zero time has no meaning here, we assume that $t > 0$ and $\tau_k > 0$ $(k = 1, 2, \cdots, n)$. Solutions to Equation (2) can be expressed as a recursive Heaviside step function with an arbitrary recursion depth *n*,

$$U^n(t, \mathbf{T}^n) = H\left[-t + \tau_n U^{n-1}(t, \mathbf{T}^{n-1})\right],$$

$$U^0(t, \mathbf{T}^0) = 1, \quad (3)$$

where $n = 1, 2, \cdots$, $\mathbf{T}^n = (\tau_n, \tau_{n-1}, \cdots, \tau_1)$, and $\mathbf{T}^0$ is a null vector. Note that Equation (3) satisfies Equation (2) only when $t \le \tau_k$ $(k = 1, 2, \cdots, n)$. The Heaviside step function (Kreyszig, 1999) is defined in Equation (4),

$$H(-t + t_0) = \begin{cases} 1, & t \le t_0 \\ 0, & t > t_0 \end{cases} \quad (4)$$

If $\tau_n$ is the minimum in $\mathbf{T}^n$, the solution has a unique shape, regardless of $\tau_k$, $(k = 1,\ 2,\ \cdots,\ n-1)$. Figure 4 shows the shape of the function.

The partial derivative of the solution of Equation (3) with respect to $t$ is:

$$\frac{\partial}{\partial t}U^n(t, \mathbf{T}^n) = \delta\left[-t + \tau_n U^{n-1}(t, \mathbf{T}^{n-1})\right]\left[-1 + \tau_n \frac{\partial}{\partial t}U^{n-1}(t, \mathbf{T}^{n-1})\right] \quad (5)$$

The first Dirac delta function in the derivatives reflects the consciousness and the following functions reflect the influences of the unconsciousness. Because an infinite number of solutions exist, an infinite number of partial derivatives also exist. Moreover, the partial derivatives cannot be uniquely defined (Bracewell 2000). Although the shapes of the solutions are the same, their derivatives are not the same.

## 5. A situation and human mental state

To postulate our theory, we start and focus on the ultimate simple case. Human mental states can be divided into the consciousness and unconsciousness. Mental states are not located in space (Davies 1983), but they are influenced by time from the past to the present. We cannot measure the size of mental states quantitatively, and each person has a unique mental state in the same situation. For example, Jane, Paul, and Bill think and feel differently or equivalently though they are watching the same movie. The model we suggest is a function of time; however, it is not a function of space. The derivative of this recursive Heaviside step function cannot be uniquely defined, and we cannot define their size. However, we can explain various mental states derived from one situation by using this model.

Table 1 indicates one example situation in our model: Jane, Paul, and Bill met a stray dog when they were walking together. Jane showed no interest in the dog because she has no meaningful experience related to a dog. Paul said that the dog is cute because he grew up with several pet dogs. On the other hand, Bill tried to stay away from the dog because a stray dog had bitten him during childhood.

Each person recognizes the situation as a corresponding function. The situation can mean $U^1$ to Jane, $U^2$ to Paul, $U^3$ to Bill, etc. The shapes of the graphs are equivalent though the functions indicating each graph are different. The mental state of a person who is experiencing the situation is represented by the derivative of the function corresponding to what he or she recognized. In this case, the mental state of Jane is:

$$\frac{\partial U^1}{\partial t} = -\delta\left(-t + \tau_1^J\right) \tag{6}$$

where $\tau_1^J$ is the time variable for Jane. The mental state of Paul is:

$$\frac{\partial U^2}{\partial t} = \delta\left[-t + \tau_2^P H\left(-t + \tau_1^P\right)\right]\left[-1 - \tau_2^P \delta\left(-t + \tau_1^P\right)\right] \tag{7}$$

where $\tau_1^P$ and $\tau_2^P$ are the time variables for Paul. Finally, the mental state of Bill is:

$$\begin{aligned}\frac{\partial U^3}{\partial t} = {} & \delta\left[-t + \tau_3^B H\left\{-t + \tau_2^B H\left(-t + \tau_1^B\right)\right\}\right] \\ & \times\left[-1 + \tau_3^B \delta\left\{-t + \tau_2^B H\left(-t + \tau_1^B\right)\right\}\left\{-1 - \tau_2^B \delta\left(-t + \tau_1^B\right)\right\}\right]\end{aligned} \tag{8}$$

where $\tau_1^B$, $\tau_2^B$, and $\tau_3^B$ are the time variables for Bill. Table 2 summarizes these functions. The minimum time variable for each one should satisfy the following conditions for the situation so that the shapes of $U^1$, $U^2$, and $U^3$ are equivalent:

$$\tau_1^P = \tau_2^J = \tau_3^B \quad (9)$$

Complex mental processes of the consciousness and unconsciousness determine a person's mental state. The derivatives are functions of time, but they are not functions of space and we cannot define the size of the derivatives. Because we cannot define magnitudes of delta functions explicitly unless we define it in terms of distribution (Roach, 1970; Stakgold, 1998), many mental states can be derived from one situation. Accordingly, we can use this model to explain human mental states in the ultimately simple incident.

## 6. Remembering and forgetting

The mental model can also be used to explain remembering and forgetting. For example, when Paul encountered a situation that is recognized by Paul as $U^3$, his mental state can be expressed as:

$$\begin{aligned}\frac{\partial U^3}{\partial t} &= \delta\left[-t+\tau_3 H\left\{-t+\tau_2 H\left(-t+\tau_1\right)\right\}\right] \\ &\quad \times\left[-1+\tau_3\delta\left\{-t+\tau_2 H\left(-t+\tau_1\right)\right\}\left\{-1-\tau_2\delta\left(-t+\tau_1\right)\right\}\right] \quad (10)\end{aligned}$$

If this partial derivative has a value at some $t$, Paul remembers something related to the situation he experienced at that time. If the partial derivative is zero for all $t$, Paul cannot remember anything related to the situation. When $t \neq \tau_3$, the first delta function in Equation (10) is zero and, therefore, the partial derivative is zero and Paul remembers nothing. Even when $t = \tau_3$, the first delta function is zero if $\tau_3 > \tau_2$ or $\tau_3 > \tau_1$. Therefore, Paul's remembering state occurs if, and only if, $t = \tau_3$, $\tau_3 \leq \tau_2$, and $\tau_3 \leq \tau_1$. When $t = \tau_3 < \tau_2 < \tau_1$ or $t = \tau_3 < \tau_1 < \tau_2$, the mental state is:

$$\frac{\partial U^3}{\partial t} = -\delta\left(-t+\tau_3\right) = -\Delta \quad (11)$$

where $\Delta = \delta(-t+\tau_3)$. When $t = \tau_3 = \tau_2 < \tau_1$, the mental state is:

$$\frac{\partial U^3}{\partial t} = \Delta(-1-\tau_3\Delta) \qquad (12)$$

When $t = \tau_3 = \tau_1 < \tau_2$, the mental state is:

$$\frac{\partial U^3}{\partial t} = -\Delta \qquad (13)$$

Finally, when $t = \tau_3 = \tau_2 = \tau_1$, Paul's mental state is:

$$\frac{\partial U^3}{\partial t} = \Delta\left[-1+\tau_3\Delta(-1-\tau_3\Delta)\right] \quad (14)$$

Therefore, many remembering states can exist for Paul according to the relative size of $t$ and $\tau_k, (k = 1,2,3)$. Because we cannot define the size of the delta function (Bracewell 2000), Paul's mental state cannot be defined explicitly; however, we can conclude that Paul can remember more things as there are more partial derivatives with a value that is not zero.

## 7. Statistical analysis

In order to observe statistical characteristics of the mental states described by our simple equation, we simulated every sequence of $\tau$s that can occur for a discrete time set of t (t=1, 2, 3, 4, and 5). We calculated the mental state $(U^5)'$ for each case, where

$$U^5(t,\mathbf{T}^5) = H\left[-t+\tau_5 H\left[-t+\tau_4 H\left[-t+\tau_3 H\left[-t+\tau_2 H\left[-t+\tau_1\right]\right]\right]\right]\right] \qquad (15)$$

The total number of cases was $5^5$ = 3,125, because each $\tau$ can have one of five discrete time values. Among them, the number of cases with non-zero mental state was 979. Figure 5 shows the number of mental states with non-zero values for each discrete time. When $t$ = 5, the mental state of only one case ($\tau_1 = \tau_2 = \tau_3 = \tau_4 = \tau_5 = 5$) is non-zero. Figure 6 shows the distribution of non-zero values according to the number of multiplications of delta functions. For example, number 3 indicates that the highest order of multiplication of delta functions in the mental state equation is 3. The value of a delta function is not defined. However, if we assume that the value is one, we can calculate the value of the mental state equation. Table 3 shows the calculated non-zero values and their frequencies.

The statistical analysis requires heavy computational resources because an evaluation of only 10,000 discrete time steps requires $10{,}000^{10{,}000}$ simulations. Even modern super computers cannot afford this analysis, which shows the extreme complexity of human mental states and the difficulty in mathematical studies of the human mind.

**8. Conclusions**

We showed that human mental states are not resolvable in terms of quantitative number by combining recursive Heaviside step function with parallel universe theory. We suggested a mathematical model that explains the various mental states of people who are experiencing the same situation. Each person's mental state is a result of complex processes between the person's consciousness and unconsciousness. A change in one's mental state makes one's universe diverge on the theory of parallel universe. Confusions in time ordering in one's universe can result in forgetting. This theory deals with extremely simple cases only. However, this work may be a new path for realizing artificial intelligence, overcoming mental disorders, and activating or understanding creativity.

We showed that human mental states cannot be expressed quantitatively using mathematics due to the extreme complexity of human mental states. Because an infinite number of time sequences are required, numerical analyses using computers are prohibitive even to the extremely advanced computers.

We cannot apply integral transforms such as Fourier transform or Laplace transform to the mental states of human minds since human mental states are mostly defined by multiple multiplication of delta functions (Zauderer, 1983). Subsequent research is needed to reveal unknown influences of the consciousness and the subdivided unconsciousness on the mind and body. We hope that future studies will provide new insight into the human mind.

## Appendix

### A. Summation and subtraction of mental states

In this paper, our prime goal is to describe human mental states using a simple equation. Human mental states are extremely complex and diverse because of the interactions between each state. Therefore, two human mental states can form another complex mental state. In this section, we show possible ways to simplify derived mental states.

Due to characteristics of the recursive Heaviside step function, operations between two functions have interesting features. For example, the derivative of a multiplication between two Heaviside step functions $U^1\left(t,\mathbf{T}^1\right)=H\left[-t+\tau_1\right]$ and $U^2\left(t,\mathbf{T}^2\right)=H\left[-t+\tau_2 H\left[-t+\tau_1\right]\right]$ can be expressed as

$$\left(U^1U^2\right)'=\left(U^1\right)'U^2+U^1\left(U^2\right)'=\left(U^1\right)'+\left(U^2\right)' \quad \text{(A1)}$$

where

$$\left(U^1\right)'=-\delta\left(-t+\tau_1\right) \quad \text{(A2)}$$

and

$$\left(U^2\right)'=\delta\left[-t+\tau_2 H\left(-t+\tau_1\right)\right]\left[-1-\tau_2\delta\left(-t+\tau_1\right)\right] \quad \text{(A3)}$$

On the other hand, the derivative of a division can be expressed as:

$$\left(\frac{U^1}{U^2}\right)'=\frac{U^2\left(U^1\right)'-U^1\left(U^2\right)'}{\left(U^2\right)^2}=\left(U^1\right)'-\left(U^2\right)' \quad \text{(A4)}$$

Therefore, a mental state with respect to a situation composed by multiplying two situations equals the sum of mental states for each situation. A mental state with respect to a situation composed by dividing two situations equals the difference between mental states for each situation.

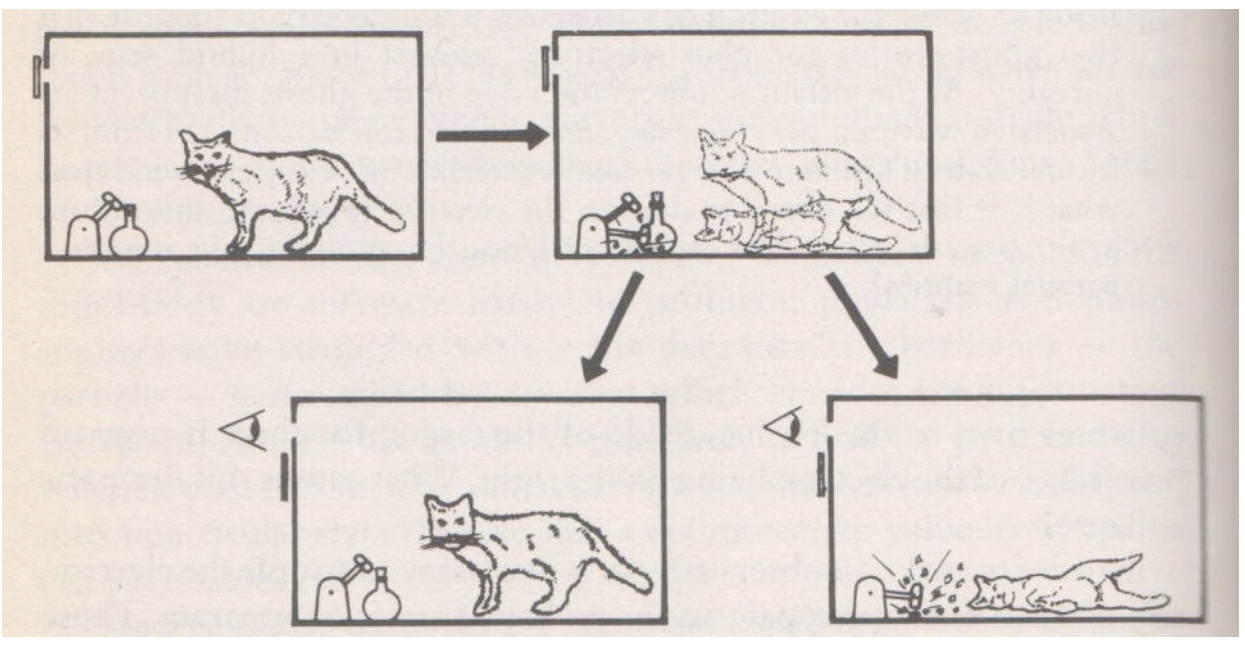

Fig. 1: The sad tale of Schrödinger's cat. A quantum process can trigger the release of cyanide with a 50:50 probability. Quantum theory requires that the system develop into a ghost-like hybrid state of live-dead cat until an observation is made, when either a live cat or a dead cat will be perceived. This thought experiment highlights the unusual implications surrounding the act of observation in the quantum theory (excerpted from (Davies, 1983)).

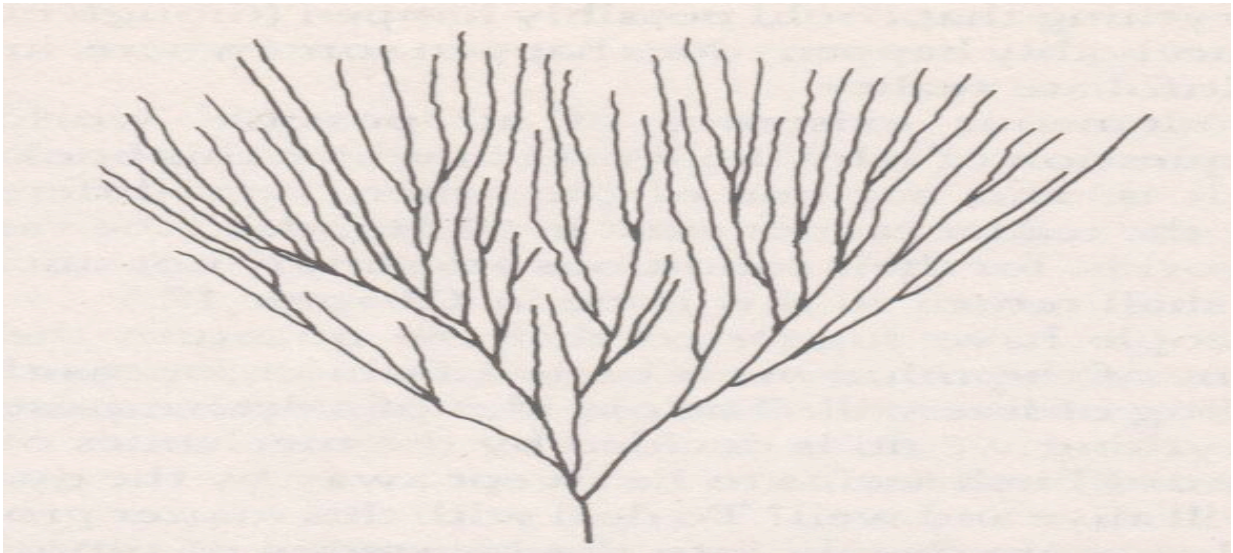

Fig. 2: Diverging time axes (excerpted from (Davies, 1983)).

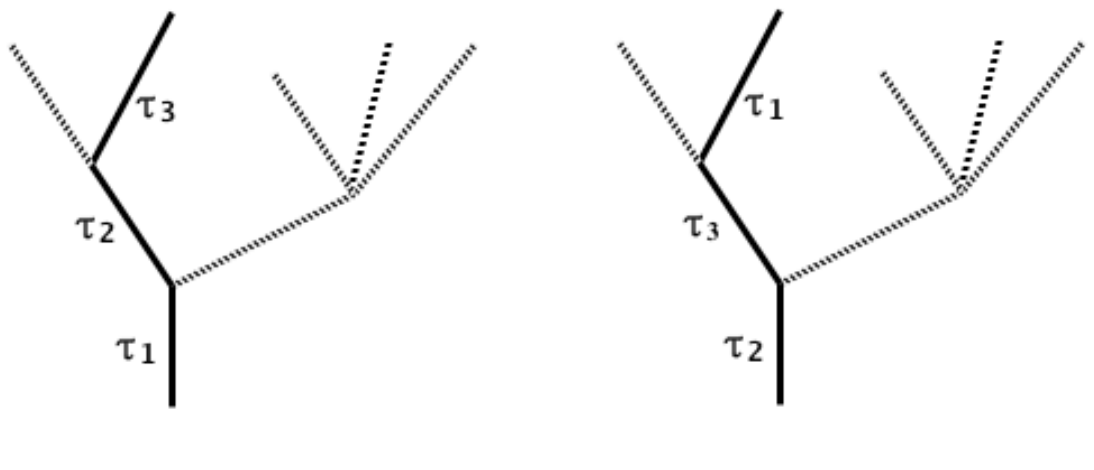

Fig. 3: Order of time axes affects one's mental state. Two of six possible cases in a universe with three time axes.

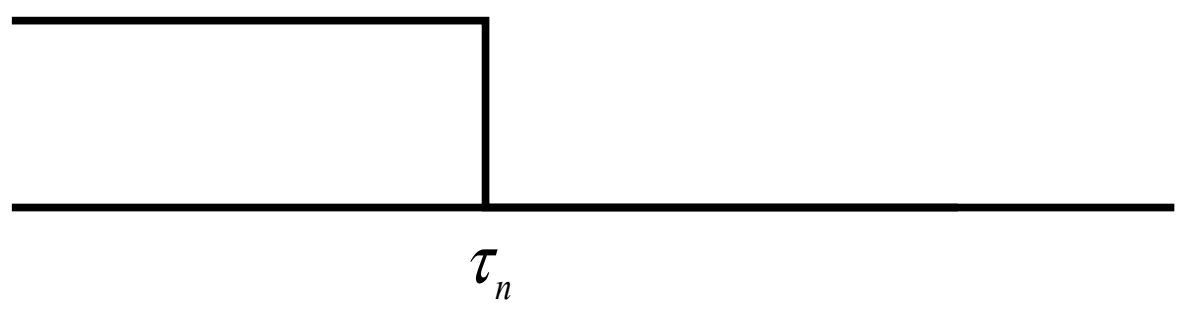

Fig. 4: The shape of the function in Equation (3).

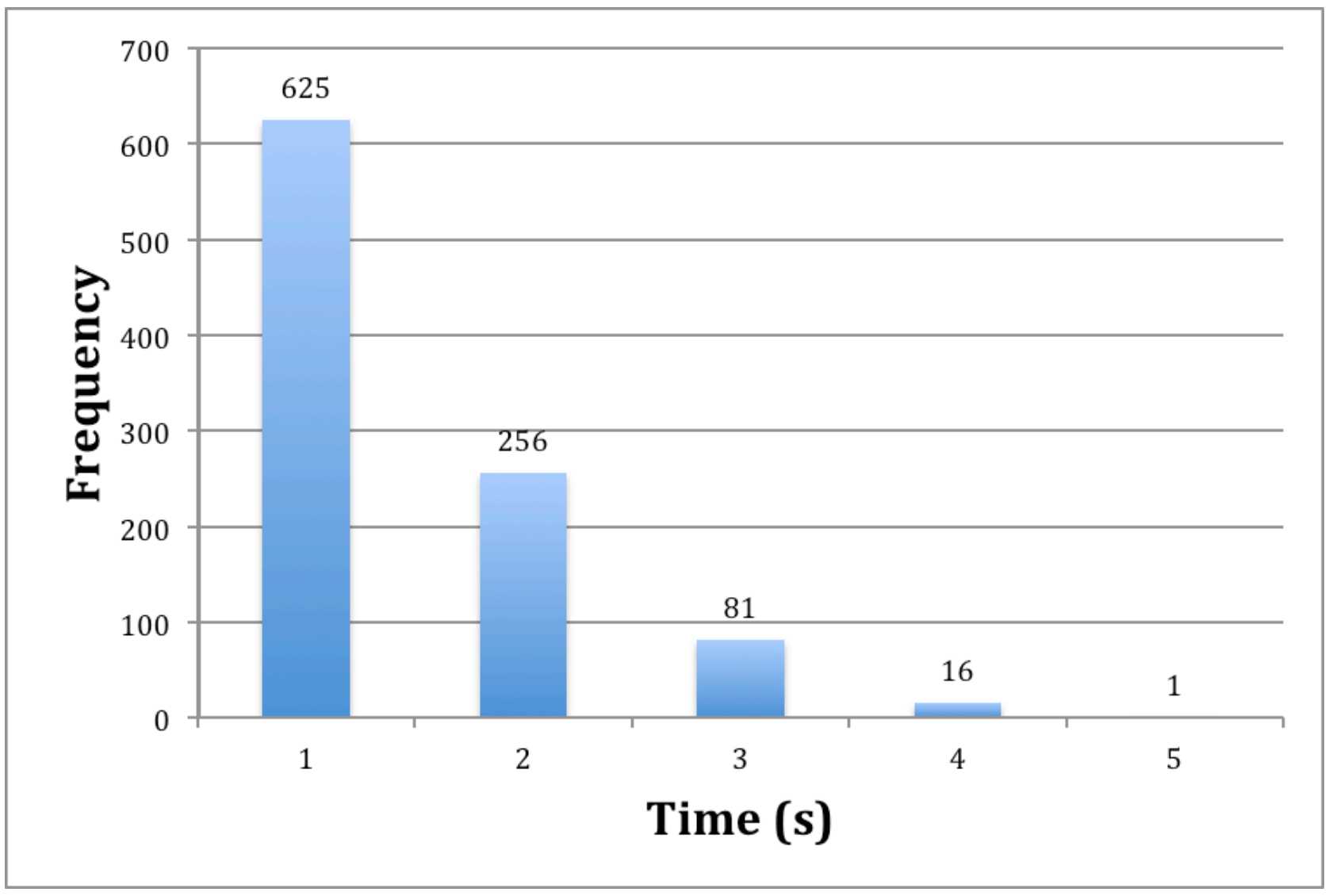

Fig. 5: The distribution of non-zero values for each discretized time.

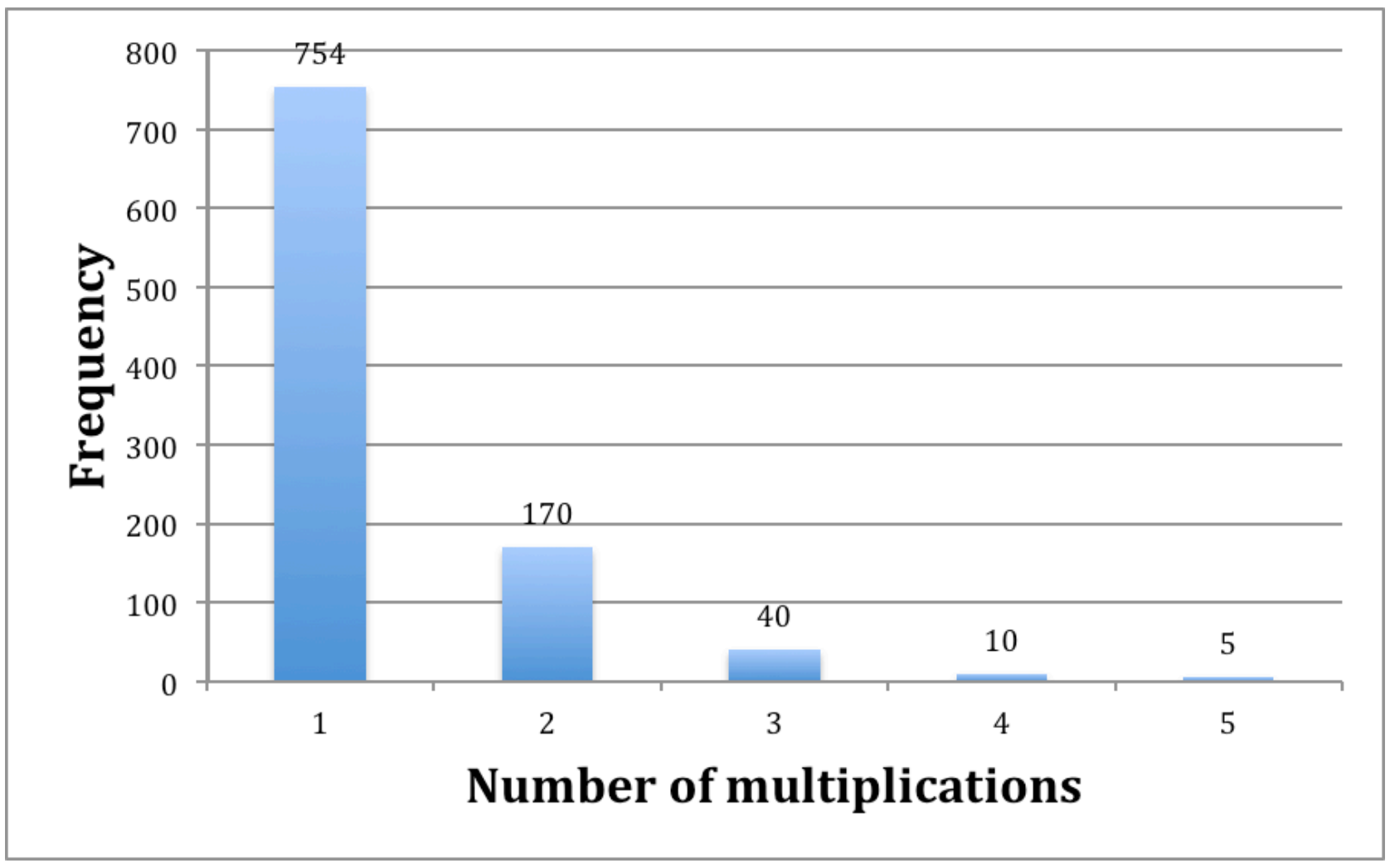

Fig. 6: The distribution of non-zero values according to the number of multiplications of delta functions.

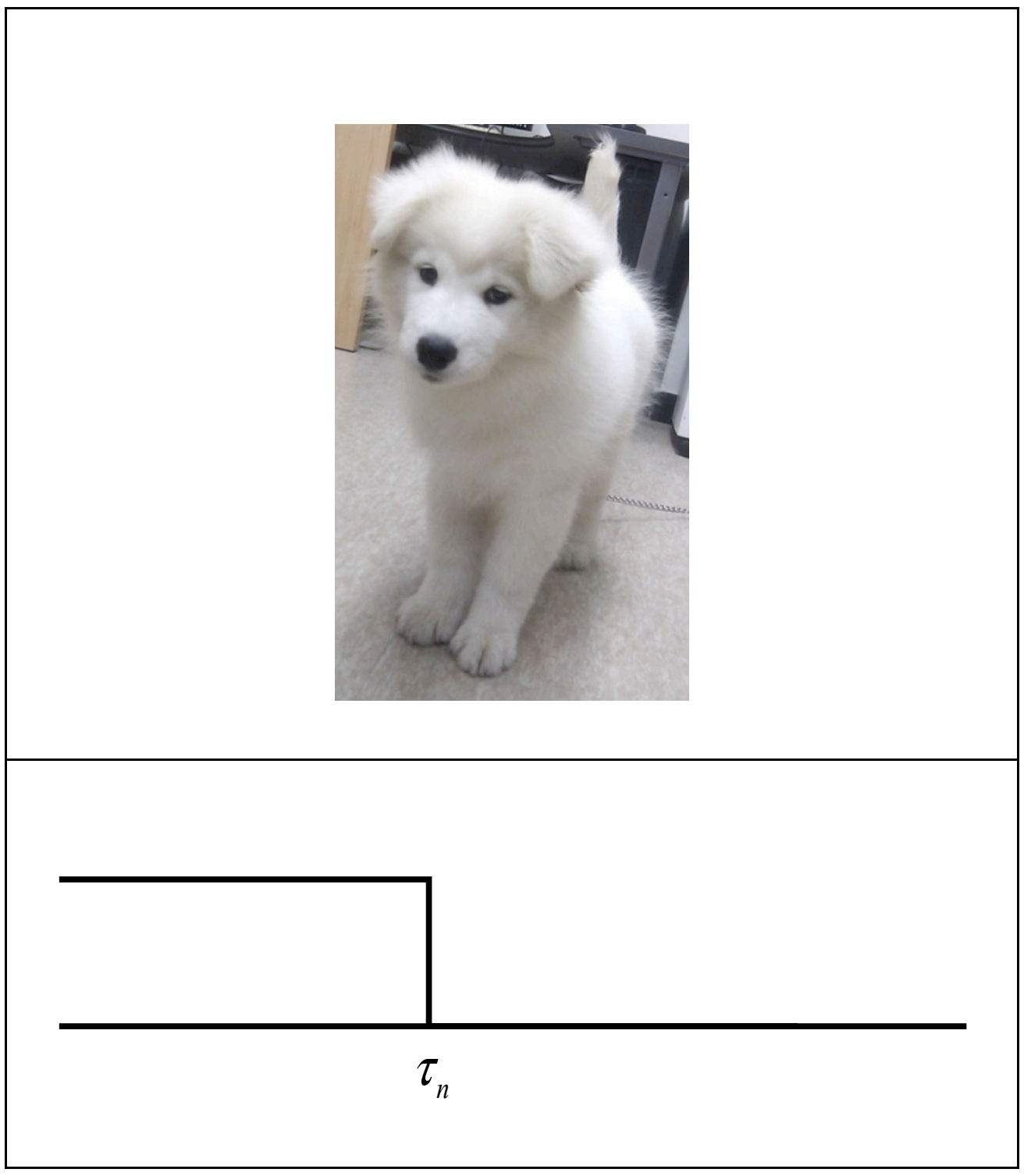

Table 1. A situation (meeting a stray dog) and the graph indicating the situation.

| Name | Mental state |
| --- | --- |
| Jane | 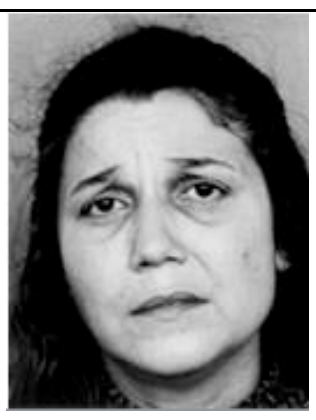 $U^1\left(t,\mathbf{T}^1\right)=H\left[-t+\tau_1^J\right]$ $\frac{\partial U^1}{\partial t}=-\delta\left(-t+\tau_1^J\right)$ |
| Paul | 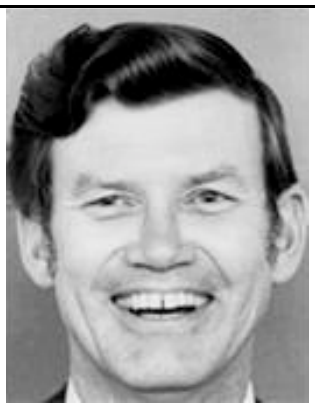 $U^2\left(t,\mathbf{T}^2\right)=H\left[-t+\tau_2^P H\left[-t+\tau_1^P\right]\right]$ $\frac{\partial U^2}{\partial t}=\delta\left[-t+\tau_2^P H\left(-t+\tau_1^P\right)\right]\left[-1-\tau_2^P\delta\left(-t+\tau_1^P\right)\right]$ |
| Bill | 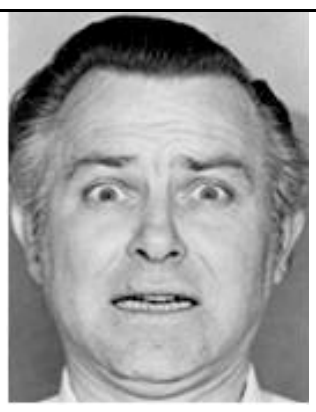 $U^3\left(t,\mathbf{T}^3\right)=H\left[-t+\tau_3^P H\left[-t+\tau_2^P H\left[-t+\tau_1^P\right]\right]\right]$ $\frac{\partial U^3}{\partial t}=\delta\left[-t+\tau_3^B H\left\{-t+\tau_2^B H\left(-t+\tau_1^B\right)\right\}\right] \times\left[-1+\tau_3^B\delta\left\{-t+\tau_2^B H\left(-t+\tau_1^B\right)\right\}\left\{-1-\tau_2^B\delta\left(-t+\tau_1^B\right)\right\}\right]$ |

Table 2. Functions indicating the situation and mental states of Jane, Paul, and Bill (Figures from http://linguistics.berkeley.edu).

| Value of mental state | Frequency |
|---|---|
| -781 | 1 |
| -341 | 1 |
| -121 | 1 |
| -85 | 1 |
| -40 | 2 |
| -31 | 1 |
| -21 | 2 |
| -15 | 3 |
| -13 | 6 |
| -7 | 12 |
| -5 | 5 |
| -4 | 22 |
| -3 | 68 |
| -2 | 100 |
| -1 | 754 |
| Total | 979 |

Table 3. The values of mental state and their frequencies.